\documentclass[11pt]{article}

\usepackage{amssymb,latexsym,amsmath,amsbsy}
\usepackage{stmaryrd}  
\usepackage{arydshln}  
\usepackage{cite}
\def\Z{{\mathbb Z}}
\def\lb{\llbracket}
\def\rb{\rrbracket}
\def\gl{\mathfrak{gl}}
\def\ssl{\mathfrak{sl}}
\def\g{\mathfrak{g}}
\def\h{\mathfrak{h}}
\def\so{\mathfrak{so}}
\def\osp{\mathfrak{osp}}
\def\pso{\mathfrak{pso}}
\def\sp{\mathfrak{sp}}
\DeclareMathOperator{\End}{End}
\DeclareMathOperator{\Str}{Str}
\DeclareMathOperator{\tr}{tr}
\DeclareMathOperator{\spn}{span}
\begin{document}

\begin{center}
{\Large \bf
$\mathbf{{\mathbb Z}_2\ \times {\mathbb Z}_2}$-graded Lie (super)algebras and generalized quantum statistics} \\[5mm] 
{\bf N.I.~Stoilova}\footnote{E-mail: stoilova@inrne.bas.bg}\\[1mm] 
Institute for Nuclear Research and Nuclear Energy, Bulgarian Academy of Sciencies,\\ 
Boul.\ Tsarigradsko Chaussee 72, 1784 Sofia, Bulgaria\\[2mm] 
{\bf J.\ Van der Jeugt}\footnote{E-mail: Joris.VanderJeugt@UGent.be}\\[1mm]
Department of  Mathematics, Computer Science and Statistics, Ghent University,\\
Krijgslaan 281-S9, B-9000 Gent, Belgium.
\end{center}


\begin{abstract}
We present systems of parabosons and parafermions in the context of Lie algebras, Lie superalgebras, 
$\Z_2\times\Z_2$-graded Lie algebras and $\Z_2\times\Z_2$-graded Lie superalgebras.
For certain relevant $\Z_2\times\Z_2$-graded Lie algebras and $\Z_2\times\Z_2$-graded Lie superalgebras,
some structure theory in terms of roots and root vectors is developed. 
The short root vectors of these algebras are identified with parastatistics operators.
For the $\Z_2\times\Z_2$-graded Lie algebra $\so_q(2n+1)$, a system consisting of two ensembles of parafermions satisfying relative paraboson relations are introduced.
For the $\Z_2\times\Z_2$-graded Lie superalgebra $\osp(1,0|2n_1,2n_2)$, a system consisting of two ensembles of parabosons satisfying relative parafermion relations are introduced.
\end{abstract}

\section{Introduction}	

Quantum physics, and in particular quantum statistics, is governed by commutators $[x,y]$ and anticommutators $\{x,y\}$ between operators $x$ and $y$. 
Hence there is a natural relation with Lie algebras and Lie superalgebras.
Given an associative algebra, the bracket $[x,y]=xy-yx$ turns it into a Lie algebra.
And for a given $\Z_2$-graded associative algebra, the bracket $\lb x, y \rb=xy - (-1)^{\xi \eta} yx$ (where $\xi$ is the degree of $x$ and $\eta$ the degree of $y$) turns it into a Lie superalgebra.
So why is it meaningful to go beyond the familiar structures of Lie algebras and Lie superalgebras, and turn to $\Z_2\times\Z_2$-graded structures?
One of the reasons comes from the Jacobi identity. 
For a Lie algebra, this reads $[x,[y,z]]+[y,[z,x]]+[z,[x,y]]=0$. 
If the Lie algebra stems from an associative algebra, these nested brackets correspond to 12 terms, canceling each other two by two. 
This trivial identity for the 12 terms can be rewritten in various ways. 
One way is, for example, $[x,\{y,z\}]+\{y,[z,x]\}-\{z,[x,y]\}=0$.
Clearly, this corresponds to the Jacobi identity for a Lie superalgebra, where $y$ and $z$ are odd elements, and $x$ is an even element.
There are many other ways to rewrite the 12 terms as nested (anti-)commutators, and most of them correspond to the Jacobi identity for a Lie superalgebra. 
But there are also other forms, for example $[x,[y,z]]+\{y,\{z,x\}\}-\{z,\{x,y\}\}=0$.
This form does not appear as the Jacobi identity for a Lie algebra or a Lie superalgebra;
it can appear only as the Jacobi identity for 
$\Z_2\times\Z_2$-graded Lie algebras or $\Z_2\times\Z_2$-graded Lie superalgebras.

Hence it is not surprising that $\Z_2\times\Z_2$-graded Lie algebras or superalgebras,
developed already in the 1970's~\cite{Rit1, Rit2, Scheunert1979},
made their appearance in mathematical physics.
In recent years, this started with the symmetries of L\'evy–Leblond equations~\cite{Aizawa1, Aizawa3}.
The $\Z_2\times\Z_2$-graded structures also appeared in graded classical mechanics~\cite{Aizawa4} and
in graded quantum mechanics~\cite{Aizawa5, Bruce2, AMD2020, Aizawa6}.
Graded superspace was investigated in~\cite{Poncin,Doi,Aizawa7,Aizawa8}, and graded bosonization in~\cite{Quesne2021}.
Some recent work focused more on the algebraic structure or on representation 
theory~\cite{Aizawa9, Aizawa10, Aizawa11, Isaac2020, Aizawa12, Aizawa13, KT2021, KT2023, Aizawa14, Lu2023, Isaac2024, Aizawa15}.

Our interest in $\Z_2\times\Z_2$-graded Lie algebras or superalgebras stems from its relation with parastatistics.
This was already recognized in~\cite{YJ,YJ2,Tolstoy2014}, and further developed in~\cite{SV1,SV2,SV3}.
In the current paper we will recall the definition of parabosons and parafermions, 
and how a mixed system leads to an orthosymplectic $\Z_2\times\Z_2$-graded Lie superalgebra.
In view of this, we investigated some classes of classical $\Z_2\times\Z_2$-graded Lie algebras~\cite{SV4, SV5} 
and Lie superalgebras~\cite{SV6}.
We will briefly recall some of the results of these papers here, but focus more on certain examples and the underlying algebraic structure.
Moreover, our examples will provide new systems of mixed parabosons and/or parafermions with an underlying 
$\Z_2\times\Z_2$-graded Lie algebra or superalgebra.

Note that another approach to parastatistics, also in the context of $\Z_2\times\Z_2$-graded Lie (super)algebras, 
has been quite successful in particular in investigating its physical consequences~\cite{Toppan2021a,Toppan2021b}.
There, the construction of multi-particle states of $\Z_2\times\Z_2$-graded paraparticles uses the braided tensor product
of Hopf algebras~\cite{Majid,Kanakoglou}.
This alternative approach to $\Z_2\times\Z_2$-graded parastatistics offers an interesting framework for 
investigating experimentally the existence of such paraparticles. 

\section{Bosons, fermions, parabosons and parafermions}

Bosons or Bose creation and annihilation operators $B_i^\pm$ (where $i$ runs over an index set) satisfy the so-called Bose-Einstein statistics, governed by the commutator relations $[B_i^-,B_j^+]=\delta_{ij}$, and all other commutators are zero.
Fermions or Fermi creation and annihilation operators $F_i^\pm$ satisfy the so-called Fermi-Dirac statistics, governed by the anti-commutator relations $\{F_i^-,F_j^+\}=\delta_{ij}$, and all other anti-commutators are zero.
It has been known for a long time that quantum theory allows for the existence of infinitely many families of paraparticles, obeying mixed-symmetry statistics.
Among the most interesting structures in generalized quantum statistics are parabosons and parafermions, introduced by 
Green~\cite{Green}.
Contrary to bosons and fermions, where the defining relations are simply commutators or anti-commutators, the defining relations for parabosons and parafermions are triple relations as nested (anti-)commutators.
A system of paraboson creation and annihilation operators $b_j^\pm$ ($j=1,\ldots,n$) is defined by the following relations:
\begin{equation}
 [\{ b_{ j}^{\xi}, b_{ k}^{\eta}\} , b_{l}^{\epsilon}]= 
 (\epsilon -\eta) \delta_{kl}b_{j}^{\xi} + (\epsilon -\xi) \delta_{jl} b_{k}^{\eta}.
\label{pb}
\end{equation}
In the right hand side of this equation, $\xi$, $\eta$ and $\epsilon$ should be treated as $\pm 1$ (corresponding to $\pm$).
Whereas a system of bosons has only one Fock space (characterized by a vacuum vector $|0\rangle$ and $B_i^- |0\rangle=0$), 
the system of parabosons has an infinite number of Fock spaces $V(p)$, each determined by a positive integer~$p$.
This Fock space is not only characterized by the relations $b_i^- |0\rangle=0$, but also by~\cite{GM} 
\begin{equation}
 \{b_j^-,b_k^+\} |0\rangle = p\,\delta_{jk}\, |0\rangle.
\end{equation}

In a similar way, a system of parafermion creation and annihilation operators $f_{ j}^\pm$ ($j=1,\ldots,m$) is defined by the following relations:
\begin{equation}
 [ [ f_{ j}^{\xi}, f_{ k}^{\eta}] , f_{l}^{\epsilon}]= 
 |\epsilon -\eta| \delta_{kl} f_{j}^{\xi} - |\epsilon -\xi| \delta_{jl} f_{k}^{\eta} .
\label{pf}
\end{equation}
There are again an infinity number of Fock spaces $W(p)$, where $p$ is a positive integer, characterized by $f_{ i}^- |0\rangle=0$ and
\begin{equation}
 [ f_{ j}^-,f_{ k}^+ ] |0\rangle = p\,\delta_{jk}\, |0\rangle.
\end{equation}
In both cases, $p$ is referred to as the order of parastatistics.
For $p=1$, the relations for parabosons (resp.\ parafermions) in the Fock space $V(1)$ (resp.\ $W(1)$) 
reduce to the relations for ordinary bosons (resp.\ fermions).

The algebraic structure underlying parafermions was soon discovered in~\cite{KT, Ryan}. 
It was shown that the Lie algebra generated by $2m$ elements $f_j^\pm$ subject to the parafermion triple relations
is the orthogonal Lie algebra $\so(2m+1)$. 
The generators $f_j^\pm$ correspond to the short (positive and negative) root vectors of $\so(2m+1)$.
The Fock space $W(p)$ is the unitary irreducible representation of $\so(2m+1)$
with lowest weight $(-\frac{p}{2}, -\frac{p}{2},\ldots, -\frac{p}{2})$ and highest weight $(\frac{p}{2}, \frac{p}{2},\ldots, \frac{p}{2})$.

Many years later it was discovered that the algebraic structure underlying parabosons is a Lie superalgebra~\cite{Ganchev} rather than a Lie algebra.
The Lie superalgebra generated by $2n$ odd elements $b_j^\pm$ subject to the paraboson triple relations
is the orthosymplectic Lie superalgebra $\osp(1|2n)$. 
In this case, the generators $b_j^\pm$ correspond to the odd (positive and negative) root vectors of $\osp(1|2n)$.
The Fock space $V(p)$ is the unitary irreducible infinite-dimensional representation of $\osp(1|2n)$
with lowest weight $(\frac{p}{2}, \frac{p}{2},\ldots, \frac{p}{2})$.

Already in the early days of parastatistics, simultaneous systems consisting of $m$ parafermions and $n$ parabosons were 
studied~\cite{GM}.
Apart from trivial ways of combining these, it was discovered that there are two non-trivial ways of combining parafermions and parabosons by means of relative triple relations~\cite{GM}.
The first of these are called the relative parafermion relations, and are determined by:
\begin{align}
&[[f_{ j}^{\xi}, f_{ k}^{\eta}], b_{l}^{\epsilon}]=0,\qquad [\{b_{ j}^{\xi}, b_{ k}^{\eta}\}, f_{l}^{\epsilon}]=0, \nonumber\\
&[[f_{ j}^{\xi}, b_{ k}^{\eta}], f_{l}^{\epsilon}]= -|\epsilon-\xi| \delta_{jl} b_k^{\eta}, \qquad
\{[f_{ j}^{\xi}, b_{ k}^{\eta}], b_{l}^{\epsilon}\}= (\epsilon-\eta) \delta_{kl} f_j^{\xi}.
\label{rel-pf}
\end{align}
The algebra generated by $2m$ even elements $f_j^\pm$ and $2n$ odd elements $b_j^\pm$ subject to the set of relations~\eqref{pb}, \eqref{pf} and~\eqref{rel-pf}
was discovered to be the Lie superalgebra $\osp(2m+1|2n)$~\cite{Palev1982}.

The second of the possible relative relations, the so-called relative paraboson relations, are given by
\begin{align}
&[[f_{ j}^{\xi}, f_{ k}^{\eta}], b_{l}^{\epsilon}]=0,\qquad
 [\{b_{ j}^{\xi}, b_{ k}^{\eta}\}, f_{l}^{\epsilon}]=0, \nonumber\\
&\{\{f_{ j}^{\xi}, b_{ k}^{\eta}\}, f_{l}^{\epsilon}\}= 
|\epsilon-\xi| \delta_{jl} b_k^{\eta}, \qquad
[\{f_{ j}^{\xi}, b_{ k}^{\eta}\}, b_{l}^{\epsilon}]= (\epsilon-\eta) \delta_{kl} f_j^{\xi}.
\label{rel-pb}
\end{align}
In this case, the algebra generated by the $2m$ elements $f_j^\pm$ and the $2n$ elements $b_j^\pm$ subject to the set of relations~\eqref{pb}, \eqref{pf} and~\eqref{rel-pb}
is not a Lie algebra nor a Lie superalgebra.
It turns out to be a $\Z_2\times\Z_2$-graded Lie superalgebra denoted by $\osp(1,2m|2n,0)$ in~\cite{Tolstoy2014}, 
or by $\pso(2m+1|2n)$ in~\cite{SV1}.

The appearance of a ``classical'' $\Z_2\times\Z_2$-graded Lie superalgebra in this parastatistics system, for which also Fock representations were studied~\cite{SV1}, led us to the investigation of classical $\Z_2\times\Z_2$-graded Lie algebras and superalgebras.
In~\cite{SV4,SV5}, four classes of $\Z_2\times\Z_2$-graded Lie algebras were constructed.
We shall not repeat this construction here, but focus on two classes in Section~3.
In particular, we develop some new structure theory for the orthogonal $\Z_2\times\Z_2$-graded Lie algebra $\so_q(2n+1)$.
The case of $\so_q(2n+1)$ is of particular interest, since it will be related to a mixed system of two ensembles of parafermions
with relative paraboson relations.

In Section~4 we reconsider the orthosymplectic $\Z_2\times\Z_2$-graded Lie superalgebra $\osp(2m_1+1,2m_2|2n_1,2n_2)$, 
defined for the first time in~\cite{SV6}.
We also develop some structure theory for this algebra.
A special case of this $\Z_2\times\Z_2$-graded Lie superalgebra will be related to a mixed system of two ensembles of parabosons
with relative parafermion relations.

\section{$\mathbf{\Z_2\times\Z_2}$-graded Lie algebras}

\subsection{Definition}

The definition of $\Z_2\times\Z_2$-graded Lie algebras goes back to~\cite{Rit1,Rit2}.
As a linear space, the $\Z_2\times\Z_2$-graded Lie algebra $\g$ is a direct sum of four subspaces:
\begin{equation}
\g=\bigoplus_{\boldsymbol{a}} \g_{\boldsymbol{a}} =
\g_{(0,0)} \oplus \g_{(0,1)} \oplus \g_{(1,0)} \oplus \g_{(1,1)} 
\label{ZZgrading}
\end{equation}
where $\boldsymbol{a}=(a_1,a_2)$ is an element of $\Z_2\times\Z_2$.
Elements of $\g_{\boldsymbol{a}}$ are denoted by $x_{\boldsymbol{a}}, y_{\boldsymbol{a}},\ldots$,
and $\boldsymbol{a}$ is called the degree, $\deg x_{\boldsymbol{a}}$, of $x_{\boldsymbol{a}}$.
Such elements are homogeneous elements.
Then $\g$ is a $\Z_2\times\Z_2$-graded Lie algebra if it admits a bilinear operation  $\lb\cdot,\cdot\rb$ which
satisfies the grading, symmetry and Jacobi identities:
\begin{align}
& \lb x_{\boldsymbol{a}}, y_{\boldsymbol{b}} \rb \in \g_{\boldsymbol{a}+\boldsymbol{b}}, \label{grading}\\
& \lb x_{\boldsymbol{a}}, y_{\boldsymbol{b}} \rb = -(-1)^{\boldsymbol{a}\cdot\boldsymbol{b}} 
\lb y_{\boldsymbol{b}}, x_{\boldsymbol{a}} \rb, \label{symmetry}\\
& \lb x_{\boldsymbol{a}}, \lb y_{\boldsymbol{b}}, z_{\boldsymbol{c}}\rb \rb =
\lb \lb x_{\boldsymbol{a}}, y_{\boldsymbol{b}}\rb , z_{\boldsymbol{c}} \rb +
(-1)^{\boldsymbol{a}\cdot\boldsymbol{b}} \lb y_{\boldsymbol{b}}, \lb x_{\boldsymbol{a}}, z_{\boldsymbol{c}}\rb \rb,
\label{jacobi}
\end{align} 
where
\begin{equation}
\boldsymbol{a}+\boldsymbol{b}=(a_1+b_1,a_2+b_2)\in \Z_2\times\Z_2, \qquad
\boldsymbol{a}\cdot\boldsymbol{b} = a_1b_2-a_2b_1.
\label{sign}
\end{equation}
By~\eqref{symmetry}, the bracket corresponds to a commutator or anti-commutator for homogeneous elements.

Note that in general a $\Z_2\times\Z_2$-graded Lie algebra is not a Lie algebra nor a Lie superalgebra, but clearly
$\g_{(0,0)}$ is a Lie subalgebra, and $\g_{(0,1)}$,  $\g_{(1,0)}$ and $\g_{(1,1)}$ are $\g_{(0,0)}$-modules.
Furthermore, $[\g_{\boldsymbol{a}}, \g_{\boldsymbol{a}}] \subset \g_{(0,0)}$ for $\boldsymbol{a}\in\Z_2\times\Z_2$,
and $\{\g_{\boldsymbol{a}}, \g_{\boldsymbol{b}}\}\subset \g_{\boldsymbol{c}}$
if $\boldsymbol{a}$, $\boldsymbol{b}$ and $\boldsymbol{c}$ are mutually distinct elements of $\{ (1,0),(0,1),(1,1)\}$.

Any associative $\Z_2\times\Z_2$-graded algebra can easily be turned into a $\Z_2\times\Z_2$-graded Lie algebra.
Indeed, let $\g$ be an associative $\Z_2\times\Z_2$-graded algebra, with a product denoted by $x\cdot y$:
\begin{equation}
\g_{\boldsymbol{a}} \cdot \g_{\boldsymbol{b}} \subset \g_{\boldsymbol{a}+\boldsymbol{b}}.
\end{equation}
Then $(\g, \lb\cdot,\cdot\rb)$ is a $\Z_2\times\Z_2$-graded Lie algebra by the bracket
\begin{equation}
\lb x_{\boldsymbol{a}} , y_{\boldsymbol{b}}\rb = x_{\boldsymbol{a}} \cdot y_{\boldsymbol{b}}- 
(-1)^{\boldsymbol{a}\cdot\boldsymbol{b}} y_{\boldsymbol{b}} \cdot x_{\boldsymbol{a}} ,
\label{bracketLA}
\end{equation}
with $\boldsymbol{a}\cdot\boldsymbol{b} = a_1b_2-a_2b_1$.

\subsection{$\mathbf{\gl_{p,q,r,s}(n)}$ and $\mathbf{\ssl_{p,q,r,s}(n)}$}

General and special linear $\Z_2\times\Z_2$-graded Lie algebras were already constructed in~\cite{Rit1,Rit2}.
Let $V$ be a $\Z_2\times \Z_2$-graded linear space of dimension $n$:
$V=V_{(0,0)} \oplus V_{(0,1)} \oplus V_{(1,0)} \oplus V_{(1,1)}$, with subspaces of dimension $p$, $q$, $r$ and $s$ respectively ($p+q+r+s=n$).
$\End(V)$ is then a $\Z_2\times \Z_2$-graded associative algebra, and turned into a 
$\Z_2\times\Z_2$-graded Lie algebra by the bracket~\eqref{bracketLA}. 
This algebra is denoted by $\gl_{p,q,r,s}(n)$.  
Its elements are of the following matrix form:
\begin{equation}
\begin{array}{c c}
    \begin{array} {@{} c c  cc @{}} \ \ p\ \ \ & \ \ \ q\ \ & \ \ \ r\ \ & \ \ s \ \ \end{array} & {} \\[-1mm]  
\left(\begin{array}{cccc} 
a_{(0,0)} & a_{(0,1)} & a_{(1,0)} & a_{(1,1)} \\ 
b_{(0,1)} & b_{(0,0)} & b_{(1,1)} & b_{(1,0)} \\ 
c_{(1,0)} & c_{(1,1)} & c_{(0,0)} & c_{(0,1)} \\ 
d_{(1,1)} & d_{(1,0)} & d_{(0,1)} & d_{(0,0)} 
\end{array}\right)
 & \hspace{0em}
		\begin{array}{l}
     p \\  q \\	r \\ s
    \end{array} \\ 
    \mbox{} 
  \end{array} \\[-12pt]  . 
\label{ZZsl}
\end{equation}	
The indices of the matrix blocks refer to the $\Z_2\times\Z_2$-grading. 
The borders of the matrix keep track of the size of the blocks.
The order of the blocks labeled by $(q,r,s)$ can be simultaneously permuted in rows and columns, giving rise to an equivalent form of $\gl_{p,q,r,s}(n)$.
Note that the defining matrices of $\gl_{p,q,r,s}(n)$ are the same as the defining matrices of $\gl(n)$, but the brackets are different.

One can check that $\tr \lb A,B \rb =0$ ($\tr$ stands for the ordinary trace), hence $\g=\ssl_{p,q,r,s}(n)$ is the subalgebra of traceless elements
of $\gl_{p,q,r,s}(n)$, and it is also a $\Z_2\times\Z_2$-graded Lie algebra.

\subsection{Subalgebras of $\mathbf{\ssl_{p,q,r,s}(n)}$}

For an element $A\in \ssl_{p,q,r,s}(n) \subset \End(V)$ with $\deg(A)=\boldsymbol{a}$, 
one can define the conjugate (or dual) $A^* \in \End(V^*)$ by the following requirement:
\begin{equation}
\langle A^* y_{\boldsymbol{b}},x\rangle= (-1)^{{\boldsymbol{a}}\cdot{\boldsymbol{b}}} \langle y_{\boldsymbol{b}}, Ax \rangle, \qquad
\forall x\in V, \quad \forall y_{\boldsymbol{b}}\in V_{\boldsymbol{b}}^*.
\label{dual}
\end{equation}
Herein $\langle\cdot,\cdot\rangle$ is natural pairing of $V$ and $V^*$.
In matrix form, this leads to the $\Z_2\times\Z_2$-graded transpose $A^T$ of $A$:
\begin{equation}
A=\left(\begin{array}{cccc} 
a_{(0,0)} & a_{(0,1)} & a_{(1,0)} & a_{(1,1)} \\ 
b_{(0,1)} & b_{(0,0)} & b_{(1,1)} & b_{(1,0)} \\ 
c_{(1,0)} & c_{(1,1)} & c_{(0,0)} & c_{(0,1)} \\ 
d_{(1,1)} & d_{(1,0)} & d_{(0,1)} & d_{(0,0)} 
    \end{array}\right), \quad
A^T=\left(\begin{array}{cccc} 
a_{(0,0)}^{\;t} & b_{(0,1)}^{\;t} & c_{(1,0)}^{\;t} & d_{(1,1)}^{\;t} \\ 
a_{(0,1)}^{\;t} & b_{(0,0)}^{\;t} & -c_{(1,1)}^{\;t} & -d_{(1,0)}^{\;t} \\ 
a_{(1,0)}^{\;t} & -b_{(1,1)}^{\;t} & c_{(0,0)}^{\;t} & -d_{(0,1)}^{\;t} \\ 
a_{(1,1)}^{\;t} & -b_{(1,0)}^{\;t} & -c_{(0,1)}^{\;t} & d_{(0,0)}^{\;t} 
    \end{array}\right),
\end{equation}
where $a^t$ is the ordinary transpose of $a$.
The following property is easy to check by means of a case by case examination of the degrees:
\begin{equation}
(AB)^T = (-1)^{{\boldsymbol{a}}\cdot{\boldsymbol{b}}} B^T A^T,
\end{equation}
where $\boldsymbol{a} = \deg A$ and $\boldsymbol{b} = \deg B$.

For ordinary Lie algebras, one can define $\so(n)$ as the subalgebra consisting of matrices $X\in\ssl(n)$ with $X^t+X=0$. 
In the current situation, one defines
\begin{equation}
\g=\so_{p,q,r,s}(n) = \{A\in \ssl_{p,q,r,s}(n) \ |\  A^T+A=0 \}.
\end{equation}
If $A,B \in \g$ are homogeneous elements, then
\begin{align}
\lb A,B\rb^T &= (AB-(-1)^{{\boldsymbol{a}}\cdot{\boldsymbol{b}}}BA)^T \nonumber\\
&=(-1)^{{\boldsymbol{a}}\cdot{\boldsymbol{b}}} B^T A^T-A^TB^T = (-1)^{{\boldsymbol{a}}\cdot{\boldsymbol{b}}} B A-AB=-\lb A,B\rb.
\end{align}
Hence $\so_{p,q,r,s}(n)$ is indeed a subalgebra of $\ssl_{p,q,r,s}(n)$, and a $\Z_2\times\Z_2$-graded Lie algebra.
The matrices $\so_{p,q,r,s}(n)$ are of the following form:
\begin{equation}
\begin{array}{c c}
    \begin{array} {@{} c c  cc @{}} \ \ p\ \ \ & \ \ q\ \ \ & \ \ r\ \ \ & \ \ s \ \ \end{array} & {} \\[-1mm]  
    \left(\begin{array}{cccc} 
a_{(0,0)} & a_{(0,1)} & a_{(1,0)} & a_{(1,1)} \\ 
-a_{(0,1)}^{\;t} & b_{(0,0)} & b_{(1,1)} & b_{(1,0)} \\ 
-a_{(1,0)}^{\;t} & b_{(1,1)}^{\;t} & c_{(0,0)} & c_{(0,1)} \\ 
-a_{(1,1)}^{\;t} & b_{(1,0)}^{\;t} & c_{(0,1)}^{\;t} & d_{(0,0)} 
    \end{array}\right) 
 & \hspace{-1mm}	\begin{array}{l}
     p \\[0mm]  q \\[0mm]	r \\[0mm] s
    \end{array} \\ 
  \end{array} 
\end{equation}
where $a_{(0,0)}$, $b_{(0,0)}$, $c_{(0,0)}$ and $d_{(0,0)}$ are antisymmetric matrices.

However, it will be useful to introduce a different orthogonal subalgebra of $\ssl_{p,q,r,s}(n)$.
The reason is the following: in the matrix definition of the Lie algebra $\so(n)$ as the set of anti-symmetric matrices $X$ (with $X^t+X=0$), 
the Cartan subalgebra does not consist of diagonal matrices.
For Lie algebras, this is no problem and for many purposes one can continue to work in this matrix form.
For the $\Z_2\times\Z_2$-graded Lie algebra $\so_{p,q,r,s}(n)$, this gives rise to complications.
To illustrate this, consider $\g=\so_{1,1,2,2}(6)$. 
The analog of the Cartan subalgebra basis of $\so(6)$ would now consist of the following elements of $\so_{1,1,2,2}(6)$:
$h_1=e_{1,4}-e_{4,1}$, $h_2=e_{2,5}+e_{5,2}$ and $h_3=e_{3,6}+e_{6,3}$, where as usual $e_{ij}$ is the matrix 
of the relevant size with zeros everywhere except a 1 on the intersection of row~$i$ and column~$j$.
Although in this example $\lb h_i, h_j \rb =0$ for $i,j \in\{1,2,3\}$, note that
\begin{equation}
h_1\in \g_{(1,0)}, \qquad h_2\in \g_{(1,0)}, \qquad h_3\in \g_{(0,1)}
\end{equation}
hence the bracket between $h_1$ and $h_3$ (or $h_2$ and $h_3$) is an anti-commutator: $[h_1,h_2]=0$, $\{h_1,h_3\}=0$, $\{h_2,h_3\}=0$.
Clearly, for the development of further structure theory, this is not desirable, and one would prefer to have a Cartan subalgebra basis consisting of commuting elements.

For this reason, it will be advantageous to introduce a different matrix form of orthogonal $\Z_2\times\Z_2$-graded Lie algebras, as has been done in~\cite{SV4, SV5}.
Recall that for classical Lie algebras of type $B$, $C$ and $D$, such matrix forms in which the Cartan subalgebra consists of diagonal matrices is well known.
For type $B$, i.e.\ the Lie algebra $\so(2n+1)$, this matrix form is given by
\begin{equation}
\begin{array}{c c}
    \begin{array} {@{} c c c @{}} \ \ n\  & \ \ n\ \ & 1\ \end{array} & {} \\[-1mm]  
    \left(
       \begin{array}{@{} c  c  c @{}}
        a & b & c\\ 
        d & -a^t & e\\
				-e^t & -c^t & 0
       \end{array}
    \right)  & \hspace{-2mm}
		\begin{array}{c}
     n \\[0.2mm]  n \\[0.2mm]	1 
    \end{array}  
   \end{array} 
\end{equation}
where the size of the block matrices is again determined by the borders, and $b$ and $d$ are anti-symmetric matrices.
The $\Z_2\times\Z_2$-graded analog of this has been constructed in~\cite{SV4, SV5}:
the $\Z_2\times\Z_2$-graded Lie algebra $\g=\so_{q}(2n+1)$, with $1\leq q\leq n-1$, consists of all matrices of the following block form:
\begin{equation}
\begin{array}{c c}
 {\begin{array} {@{} c c c c c @{}} \ \ q\ \ & \ \ \ n-q\ \ & \ \ q \ \ \ & \ \ n-q\ \ & \ 1 \ \ \end{array} } & {} \\
 \left(\begin{array}{@{} cc:cc:c @{}} a_{(0,0)}&a_{(1,1)}&b_{(0,0)}&b_{(1,1)}&c_{(0,1)}  \\[1mm]
       \tilde{a}_{(1,1)}&\tilde{a}_{(0,0)}&b_{(1,1)}^{\;t}&\tilde{b}_{(0,0)}&c_{(1,0)}  \\[1mm] \hdashline &&&&\\[-3mm]
       d_{(0,0)}&d_{(1,1)}&-a_{(0,0)}^{\;t}&\tilde{a}_{(1,1)}^{\;t}&e_{(0,1)}  \\[1mm]
       d_{(1,1)}^{\;t}&\tilde{d}_{(0,0)}&a_{(1,1)}^{\;t}&-\tilde{a}_{(0,0)}^{\;t}&e_{(1,0)}  \\[1mm] \hdashline &&&&\\[-3mm]
      -e_{(0,1)}^{\;t}&-e_{(1,0)}^{\;t}&-c_{(0,1)}^{\;t}&-c_{(1,0)}^{\;t}&0  
      \end{array}\right)
 & \hspace{-2mm}
		{\begin{array}{c} q \\[1.5mm] n-q \\[1.5mm] q \\[1.5mm] n-q \\[2mm] 1 \end{array} }
\end{array} 
\label{soB}
\end{equation}
where $b_{(0,0)}$, $\tilde{b}_{(0,0)}$, $d_{(0,0)}$ and $\tilde{d}_{(0,0)}$ are anti-symmetric matrices.
There exist some other matrix forms for $\g=\so_{q}(2n+1)$, see~\cite{SV4}.
Contrary to the $\gl$-case, the defining matrices for $\so_q(2n+1)$ and $\so(2n+1)$ are not the same.
In~\eqref{soB}, the four blocks $a_{(1,1)}^{\;t}$, $\tilde{a}_{(1,1)}^{\;t}$, $b_{(1,1)}^{\;t}$ and $d_{(1,1)}^{\;t}$ would appear with opposite signs in
the defining matrices of $\so(2n+1)$.
Note that
\begin{align*}
&\dim \g_{(0,0)}= 2n^2-n-4q(n-q)^2\\
&\dim\g_{(0,1)} = 2q, \quad \dim\g_{(1,0)} = 2(n-q)\\ 
&\dim\g_{(1,1)} = 4q(n-q),
\end{align*}
hence the dimension of $\so_{q}(2n+1)$ is the same as that of $\so(2n+1)$.
 
The elements of $\g=\so_{q}(2n+1)$ can also be characterized in a different way: 
$\g$ consists of all matrices $A$ of $\ssl_{2q,1,0,2n-2q}(2n+1)$ 
that satisfy
\begin{equation}
 A^T K +  K A=0
\label{soK}
\end{equation}
where
\begin{equation}
K= \begin{array}{c c}
\left(\begin{array}{@{} cc:cc:c @{}} 
  \ 0\  & \ 0\  & \ I\  & \ 0\ & \ 0\  \\[1mm]
  0 & 0 & 0 & -I & 0 \\[1mm] \hdashline
  I & 0 & 0 & 0  & 0 \\[1mm]
  0 & -I & 0 & 0 & 0 \\[1mm] \hdashline
	0 & 0 & 0 & 0 & 1
\end{array}\right)
 & \hspace{-2mm}
		{\begin{array}{c}      q \\[1mm] n-q \\[1mm] q \\[1mm] n-q \\[1mm] 1  \end{array} }
  \end{array} 
\end{equation}
and $I$ is the identity matrix of appropriate size.
Note that $K^T=K$ and $K^{-1}=K^t$. 
It is easy to show that if $A$ and $B$ satisfy~\eqref{soK}, then also $\lb A,B\rb$ satisfies~\eqref{soK},
thus confirming the alternative characterization of $\so_{q}(2n+1)$.

The advantage of this matrix form~\eqref{soB} is that the definition of a Cartan subalgebra is straightforward, as it consists of the set of diagonal matrices.
For $\g=\so_{q}(2n+1)$, a basis for the Cartan subalgebra $\h$ is given by
\begin{equation}
h_i=e_{i,i}-e_{n+i,n+i} \qquad i=1,\ldots,n.
\end{equation}
Now $\h\subset\g_{(0,0)}$, i.e.\ the Cartan subalgebra is just the Cartan subalgebra of the Lie algebra $\g_{(0,0)}$, which makes further structure theory feasible.
In terms of the dual basis $\epsilon_j$ ($j=1,\ldots,n$) of $\h^*$ the roots and corresponding root vectorss of $\so_q(2n+1)$ 
are given by:
\begin{equation}
\begin{array}{lllllll}
\hbox{root} && \deg && \hbox{root vector} && \\
\epsilon_j && (0,1) &\quad & e_{j,2n+1}-e_{2n+1,j+n} &\quad &  j=1,\ldots ,q \\			
\epsilon_j && (1,0) && e_{j,2n+1}-e_{2n+1,j+n} &&  j=q+1,\ldots ,n \\		
-\epsilon_j && (0,1) && e_{n+j,2n+1}-e_{2n+1,j} && j=1,\ldots ,q \\		
-\epsilon_j && (1,0) && e_{n+j,2n+1}-e_{2n+1,j} && j=q+1,\ldots ,n \\		
\epsilon_j -\epsilon_k && (0,0) && e_{jk}-e_{k+n,j+n} && j\neq k=1,\ldots,q \hbox{ or } j\neq k=q+1,\ldots,n \\		
\epsilon_j -\epsilon_k && (1,1) && e_{jk}+e_{k+n,j+n} && j=1,\ldots,q;\ k=q+1,\ldots,n \hbox{ or }\\	
 &&&&&& j=q+1,\ldots,n;\ k=1,\ldots,q \\  		
\epsilon_j +\epsilon_k && (0,0) && e_{j,k+n}-e_{k,j+n} &&  j<k=1,\ldots,q \hbox{ or } j<k=q+1,\ldots,n \\		
\epsilon_j +\epsilon_k && (1,1) && e_{j,k+n}+e_{k,j+n} && j=1,\ldots,q;\ k=q+1,\ldots,n\\  
-\epsilon_j -\epsilon_k && (0,0) && e_{j+n,k}-e_{k+n,j} &&  j<k=1,\ldots,q \hbox{ or } j<k=q+1,\ldots,n \\		
-\epsilon_j -\epsilon_k && (1,1) && e_{j+n,k}+e_{k+n,j} && j=1,\ldots,q;\ k=q+1,\ldots,n 	 
\end{array}
\label{roots}
\end{equation}
The positive roots are given by 
\begin{equation}
\Delta^+=\{ \epsilon_j \ (j=1,\ldots ,n); \epsilon_j -\epsilon_k,
\epsilon_j +\epsilon_k \ (1\leq j<k\leq n)\}
\end{equation}
but note that there are four different types of roots, according to the $\Z_2\times\Z_2$ degree:
\begin{align*}
\Delta^+_{(0,1)} &= \{ \epsilon_j \ (j=1,\ldots,q)\} \\
\Delta^+_{(1,0)} &= \{ \epsilon_j \ (j=q+1,\ldots,n)\} \\
\Delta^+_{(0,0)} &= \{ \epsilon_j - \epsilon_k, \epsilon_j+\epsilon_k \ (j<k=1,\ldots,q \hbox{ or } j<k=q+1,\ldots,n)\} \\
\Delta^+_{(1,1)} &= \{ \epsilon_j - \epsilon_k, \epsilon_j+\epsilon_k \ (j=1,\ldots,q; k=q+1,\ldots,n)\} 
\end{align*}
A set of simple roots (with their degrees) is given by
\begin{equation}
\begin{array}{cccccccc}
\epsilon_1-\epsilon_2 & \ldots & \epsilon_{q-1}-\epsilon_q & \epsilon_{q}-\epsilon_{q+1} & \epsilon_{q+1}-\epsilon_{q+2} & \ldots & \epsilon_{n-1}-\epsilon_{n} & \epsilon_n \\
(0,0) & \ldots & (0,0) & (1,1) & (0,0) & \ldots & (0,0) & (1,0)
\end{array}
\end{equation}
In other words, for the $\Z_2\times\Z_2$-graded Lie algebra $\so_q(2n+1)$ we have the same root space decomposition as for the Lie algebra $\so(2n+1)$, 
the main difference being the degree of the roots, and the fact that both commutators and anti-commutators appear among the brackets between root vectors.

In this contribution, we have concentrated on the $\Z_2\times\Z_2$-graded Lie algebra of type $B$.
Appropriate matrix forms for $\Z_2\times\Z_2$-graded Lie algebra of type $C$ and $D$ 
(analogs of the Lie algebras $\sp(2n)$ and $\so(2n)$) have been given in~\cite{SV4, SV5}  and also allow a similar structure analysis.
 
\subsection{$\mathbf{\Z_2\times\Z_2}$-graded parafermions}

Just as for $\so(2n+1)$, one can generate $\g=\so_q(2n+1)$ by means of the root vectors corresponding to the short roots.
Let us consider (the factor $\sqrt{2}$ is for convenience, to make the identification with ordinary parafermions):
\begin{equation}
f_{k}^-= \sqrt{2}(e_{j, 2n+1}-e_{2n+1,n+j}), \quad
f_{k}^+= \sqrt{2}(e_{2n+1, j}-e_{n+j,2n+1}) \qquad (k=1,\ldots , n)
\label{f-as-e}
\end{equation}
In terms of these generators, the relevant subspaces of $\g$ are given by:
\begin{align*}
& \g_{(0,1)}=\spn \{ f_{k}^{\pm},\; k=1,\ldots,q \} \\
& \g_{(1,0)}=\spn \{f_{k}^{\pm},\; k=q+1,\ldots,n  \} \\
& \g_{(0,0)}=\spn \{ [ f_{k}^\xi,  f_{l}^\eta],\; \xi, \eta =\pm, \; k,l=1,\ldots,q \; \hbox{and} \; k,l=q+1,\ldots,n   \} \\
& \g_{(1,1)}=\spn \{ \{f_{k}^{\xi}, f_{l}^{\eta}\}, \; \xi, \eta =\pm, \;k= 1,\ldots ,q,\; l=q+1,\ldots n \}.
\end{align*}
Thus we are dealing with two ensembles of parafermions, those of degree $(0,1)$ and those of degree $(1,0)$.
Within one of these ensembles, the common parafermion relations~\eqref{pf} are valid:
\begin{align}
& [[f_{ j}^{\xi}, f_{ k}^{\eta}], f_{l}^{\epsilon}]=|\epsilon -\eta|\delta_{kl} f_{j}^{\xi} - |\epsilon -\xi|\delta_{jl}f_{k}^{\eta},\nonumber \\
& \hbox{either }j,k,l=1,\ldots,q \hbox{ or else }j,k,l=q+1,\ldots,n.
\end{align}
But the ``relative commutation relations'' between the two ensembles of parafermions are of the following type: 
\begin{align}
& \{\{ f_{ j}^{\xi}, f_{ k}^{\eta}\} , f_{l}^{\epsilon}\} = |\epsilon -\eta|\delta_{kl} f_{j}^{\xi} + |\epsilon -\xi|\delta_{jl}f_{k}^{\eta}, \nonumber \\
& \hbox{either }j=1,\ldots,q,\ k=q+1,\ldots,n \hbox{ or else } j=q+1,\ldots,n,\ k=1,\ldots,q;\nonumber\\
& l=1,\ldots,n.
\label{relpb-2}
\end{align}
The situation is similar to the one described in Section~2, where a simultaneous system of parabosons and parafermions exists with so-called relative paraboson relations,
for which the underlying algebraic structure is a $\Z_2\times\Z_2$-graded Lie superalgebra.
Here, we are dealing with a simultaneous system of two ensembles of parafermions with ``relative paraboson relations''~\eqref{relpb-2},
for which the underlying algebraic structure is the $\Z_2\times\Z_2$-graded Lie algebra $\so_q(2n+1)$.

\section{$\mathbf{\Z_2\times\Z_2}$-graded Lie superalgebras}

\subsection{Definition}

The definition of a $\Z_2\times\Z_2$-graded Lie superalgebra is the same as in subsection~3.1, equations~\eqref{ZZgrading}-\eqref{sign}, 
except that the dot product in~\eqref{sign} should be changed to
\begin{equation}
\boldsymbol{a}\cdot\boldsymbol{b} = a_1b_1+a_2b_2.
\label{signLSA}
\end{equation}
In general a $\Z_2\times\Z_2$-graded Lie superalgebra is not a Lie superalgebra, since the bracket properties are different.

Just as for $\Z_2\times\Z_2$-graded Lie algebras, any $\Z_2\times\Z_2$-graded associative algebra $\g$, with a product denoted by $x\cdot y$, 
can be lifted to a $\Z_2\times\Z_2$-graded Lie superalgebra by the bracket
\begin{equation}
\lb x_{\boldsymbol{a}} , y_{\boldsymbol{b}}\rb = x_{\boldsymbol{a}} \cdot y_{\boldsymbol{b}}- 
(-1)^{\boldsymbol{a}\cdot\boldsymbol{b}} y_{\boldsymbol{b}} \cdot x_{\boldsymbol{a}}\ ,
\label{ZZbracket2}
\end{equation}
where the sign is now given by~\eqref{signLSA}.

\subsection{$\mathbf{\gl(m_1,m_2|n_1,n_2)}$ and $\mathbf{\ssl(m_1,m_2|n_1,n_2)}$}

The construction is very similar to the one in subsection~3.1.
Let $V$ be a $\Z_2\times \Z_2$-graded linear space, 
$V=V_{(0,0)} \oplus V_{(1,1)} \oplus V_{(1,0)} \oplus V_{(0,1)}$,
with subspaces of dimension $m_1,m_2,n_1$ and $n_2$ respectively.
$\End(V)$ is then a $\Z_2\times \Z_2$-graded associative algebra. 
By the bracket~\eqref{ZZbracket2} this is turned into a $\Z_2\times\Z_2$-graded Lie superalgebra, denoted by $\gl(m_1,m_2|n_1,n_2)$.
In matrix block form, the elements are given by 
\begin{equation}
A = \begin{array}{c c}
    {\begin{array} {@{} c c  cc @{}}  m_1\ & \ m_2\ & \ n_1\ & \ \ n_2 \ \end{array} } & {} \\  
    \left(\begin{array}{cccc} 
a_{(0,0)} & a_{(1,1)} & a_{(1,0)} & a_{(0,1)} \\[1mm] 
b_{(1,1)} & b_{(0,0)} & b_{(0,1)} & b_{(1,0)} \\[1mm] 
c_{(1,0)} & c_{(0,1)} & c_{(0,0)} & c_{(1,1)} \\[1mm] 
d_{(0,1)} & d_{(1,0)} & d_{(1,1)} & d_{(0,0)} 
    \end{array}\right)
 & \hspace{-2mm} 
		{\begin{array}{l}
     m_1 \\[1mm]  m_2 \\[1mm]	n_1 \\[1mm] n_2
    \end{array} } \\ 
  \end{array} \quad . 
\end{equation}
As before, the indices of a block refer to the $\Z_2\times\Z_2$-grading and the size of the blocks are indicated in the border.
Clearly, the set of matrices of the Lie algebra $\gl(m_1+m_2+n_1+n_2)$, of the Lie superalgebra $\gl(m_1+m_2|n_1+n_2)$, 
of the $\Z_2\times\Z_2$-graded Lie algebra $\gl_{m_1,m_2,n_1,n_2}(m_1+m_2+n_1+n_2)$ 
and of the $\Z_2\times\Z_2$-graded Lie superalgebra $\gl(m_1,m_2|n_1,n_2)$ are all the same, but of course the bracket is different in all of these cases.

It is easy to check that $\Str \lb A,B \rb =0$, where $\Str(A)=\tr(a_{(0,0)})+\tr(b_{(0,0)})-\tr(c_{(0,0)})-\tr(d_{(0,0)})$ 
is the graded supertrace in terms of the ordinary trace $\tr$.
Hence $\ssl(m_1,m_2|n_1,n_2)$ is defined as the subalgebra of elements of $\gl(m_1,m_2|n_1,n_2)$ with graded supertrace equal to~0.

\subsection{Subalgebras of $\mathbf{\ssl(m_1,m_2|n_1,n_2)}$}

For an element $A\in \ssl(m_1,m_2|n_1,n_2) \subset \End(V)$ with $\deg(A)=\boldsymbol{a}$, 
the conjugate $A^* \in \End(V^*)$ is defined by~\eqref{dual}, except that now the sign in $(-1)^{{\boldsymbol{a}}\cdot{\boldsymbol{b}}}$ 
is determined by~\eqref{signLSA}.
In matrix form, this yields the $\Z_2\times\Z_2$-graded supertranspose $A^{ST}$ of $A$:
\begin{equation}
A=\left(\begin{array}{cccc} 
a_{(0,0)} & a_{(1,1)} & a_{(1,0)} & a_{(0,1)} \\ 
b_{(1,1)} & b_{(0,0)} & b_{(0,1)} & b_{(1,0)} \\ 
c_{(1,0)} & c_{(0,1)} & c_{(0,0)} & c_{(1,1)} \\ 
d_{(0,1)} & d_{(1,0)} & d_{(1,1)} & d_{(0,0)} 
    \end{array}\right), \qquad
A^{ST}=\left(\begin{array}{cccc} 
a_{(0,0)}^{\;t} & b_{(1,1)}^{\;t} & -c_{(1,0)}^{\;t} & -d_{(0,1)}^{\;t} \\ 
a_{(1,1)}^{\;t} & b_{(0,0)}^{\;t} & c_{(0,1)}^{\;t} & d_{(1,0)}^{\;t} \\ 
a_{(1,0)}^{\;t} & -b_{(0,1)}^{\;t} & c_{(0,0)}^{\;t} & -d_{(1,1)}^{\;t} \\ 
a_{(0,1)}^{\;t} & -b_{(1,0)}^{\;t} & -c_{(1,1)}^{\;t} & d_{(0,0)}^{\;t} 
   \end{array}\right).
\end{equation}
The main property is again:
\begin{equation}
(AB)^{ST} = (-1)^{{\boldsymbol{a}}\cdot{\boldsymbol{b}}} B^{ST} A^{ST}
\end{equation}
with the sign determined by~\eqref{signLSA}.

The orthosymplectic $\Z_2\times\Z_2$-graded Lie superalgebra $\osp(2m_1+1,2m_2|2n_1,2n_2)$ can be defined as
the set of matrices $A$ from $\ssl(2m_1+1,2m_2|2n_1,2n_2)$ such that
\begin{equation}
A^{ST} J + JA=0
\end{equation}
where
\begin{equation}
J =
\left(\begin{array}{ccccc} 
0 & I_{m_1+m_2} & 0 & 0 & 0 \\ 
I_{m_1+m_2} & 0 & 0 & 0 & 0 \\ 
0 & 0 & 1 & 0 & 0 \\ 
0 & 0 & 0 & 0 & I_{n_1+n_2} \\ 
0 & 0 & 0 & -I_{n_1+n_2} & 0 
\end{array}\right).
\end{equation}	

The general matrix form of $\osp(2m_1+1,2m_2|2n_1,2n_2)$ has been given in~\cite{SV6}, and we do not repeat it here, 
but we shall study the structure of the special case $\g=\osp(1,0|2n_1,2n_2)$, because of its significance for parastatistics.
The defining matrices of $\g$ are
\begin{equation}
\begin{array}{c c}
 {\begin{array} {@{} c c c c c @{}} \ 1 \ \ \ & \ \ \ n_1\ \ & \ \ n_2 \ \ \ & \ \ n_1\ \ & \ \ n_2 \ \ \end{array} } & {} \\
 \left(\begin{array}{@{} ccccc @{}} 
       0 & a_{(1,0)} & a_{(0,1)} & \tilde{a}_{(1,0)} & \tilde{a}_{(0,1)}   \\[1mm]
       \tilde{a}_{(1,0)}^{\;t} & b_{(0,0)} & b_{(1,1)} & \tilde{b}_{(0,0)} & \tilde{b}_{(1,1)}  \\[1mm] 
       \tilde{a}_{(0,1)}^{\;t} & c_{(1,1)} & c_{(0,0)} & -\tilde{b}_{(1,1)}^{\;t} & \tilde{c}_{(0,0)}  \\[1mm]
       -a_{(1,0)}^{\;t} & \tilde{d}_{(0,0)} & d_{(1,1)} & -b_{(0,0)}^{\;t} & c_{(1,1)}^{\;t}  \\[1mm] 
       -a_{(0,1)}^{\;t} & -d_{(1,1)}^{;t} & \tilde{d}_{(0,0)} & b_{(1,1)}^{\;t} & -c_{(0,0)}^{\;t}  
      \end{array}\right)
 & \hspace{-1mm}
		{\begin{array}{c} 1 \\[1mm] n_1 \\[1mm] n_2 \\[1mm] n_1 \\[1mm] n_2 \end{array} }
\end{array} ,
\label{osp}
\end{equation}
where $\tilde{b}_{(0,0)}$,  $\tilde{c}_{(0,0)}$,  $d_{(0,0)}$ and $\tilde{d}_{(0,0)}$ are symmetric matrices.

The matrix form of $\g$ is such that the definition of a Cartan subalgebra $\h$ is easy and consists of the set of diagonal matrices.
A basis for $\h$ is given by
\begin{equation}
h_i=e_{i+1,i+1}-e_{n_1+n_2+i+1,n_1+n_2+i+1}, \qquad i=1,\ldots,n_1+n_2.
\end{equation}
The Cartan subalgebra is then a subalgebra of the Lie algebra $\g_{(0,0)}$.
In terms of the dual basis $\delta_i$ ($i=1,\ldots,n_1+n_2$) of $\h^*$ the root vectors and corresponding roots of $\g$ 
can be determined.
The short roots and root vectors are given by
\begin{equation}
\begin{array}{lllllll}
\hbox{root} && \deg && \hbox{root vector} && \\
-\delta_j &\quad & (1,0) &\quad & e_{1,j+1}-e_{n_1+n_2+j+1,1} &\quad &  j=1,\ldots ,n_1 \\			
-\delta_j &\quad & (0,1) &\quad & e_{1,j+1}-e_{n_1+n_2+j+1,1} &\quad &  j=n_1+1,\ldots ,n_1+n_2 \\			
\delta_j &\quad & (1,0) &\quad & e_{1,n_1+n_2+j+1}+e_{j+1,1} &\quad &  j=1,\ldots ,n_1 \\			
\delta_j &\quad & (0,1) &\quad & e_{1,n_1+n_2+j+1}+e_{j+1,1} &\quad &  j=n_1+1,\ldots ,n_1+n_2 			
\label{roots1}
\end{array}
\end{equation}
The long roots (and root vectors) are 
\begin{equation}
\begin{array}{lllll}
\hbox{root} && \deg &&\hbox{root vector}   \\
\delta_j -\delta_k && (0,0) &\quad & e_{j+1,k+1}-e_{n_1+n_2+k+1,n_1+n_2+j+1}  \\
  &&&&\qquad j\neq k=1,\ldots,n_1 \hbox{ or } j\neq k=n_1+1,\ldots,n_1+n_2 \\		
\delta_j -\delta_k && (1,1) && e_{j+1,k+1}+e_{n_1+n_2+k+1,n_1+n_2+j+1}  \\
  &&&&\qquad j=1,\ldots,n_1;\ k=n_1+1,\ldots,n_1+n_2 \hbox{ or } \\
	&&&&\qquad j=n_1+1,\ldots,n_1+n_2;\ k=1,\ldots,n_1 \\  		
\delta_j +\delta_k && (0,0) && e_{j+1,n_1+n_2+k+1}+e_{k+1,n_1+n_2+j+1}  \\
  &&&&\qquad j\leq k=1,\ldots,n_1 \hbox{ or } j\leq k=n_1+1,\ldots,n_1+n_2 \\		
\delta_j +\delta_k && (1,1) && e_{j+1,n_1+n_2+k+1}-e_{k+1,n_1+n_2+j+1} \\
  &&&&\qquad j=1,\ldots,n_1;\ k=n_1+1,\ldots,n_1+n_2 \\  
-\delta_j -\delta_k && (0,0) && e_{n_1+n_2+j+1,k+1}+e_{n_1+n_2+k+1,j+1}  \\
  &&&&\qquad j\leq k=1,\ldots,n_1 \hbox{ or } j \leq k=n_1+1,\ldots,n_1+n_2 \\		
-\delta_j -\delta_k && (1,1) && e_{n_1+n_2+j+1,k+1}-e_{n_1+n_2+k+1,j+1}  \\
  &&&&\qquad j=1,\ldots,n_1;\ k=n_1+1,\ldots,n_1+n_2 	 
\end{array}
\label{roots2}
\end{equation}

The positive roots are given by 
\begin{align}
\Delta^+ &=\{ \delta_j \ (j=1,\ldots ,n_1+n_2); \delta_j -\delta_k  \ (1\leq j<k\leq n_1+n_2); \nonumber\\
& \delta_j +\delta_k \ (1\leq j\leq k\leq n_1+n_2)\},
\end{align}
and there are again four different types of roots, according to the $\Z_2\times\Z_2$ degree:
\begin{align*}
\Delta^+_{(1,0)} &= \{ \delta_j \ (j=1,\ldots,n_1)\} \\
\Delta^+_{(0,1)} &= \{ \delta_j \ (j=n_1+1,\ldots,n_1+n_2)\} \\
\Delta^+_{(0,0)} &= \{ \delta_j - \delta_k \ (j<k=1,\ldots,n_1 \hbox{ or } j<k=n_1+1,\ldots,n_1+n_2),  \\
                 &   \delta_j+\delta_k \ (j\leq k=1,\ldots,n_1 \hbox{ or } j\leq k=n_1+1,\ldots,n_1+n_2) \} \\
\Delta^+_{(1,1)} &= \{ \delta_j - \delta_k, \delta_j+\delta_k \ (j=1,\ldots,n_1; k=n_1+1,\ldots,n_1+n_2)\} 
\end{align*}
A set of simple roots (with their degrees) is now given by
\begin{equation}
\begin{array}{cccccccc}
\delta_1-\delta_2 & \ldots & \delta_{n_1-1}-\delta_{n_1} & \delta_{n_1}-\delta_{n_1+1} & \delta_{n_1+1}-\delta_{n_1+2} & \ldots & \delta_{n_1+n_2-1}-\delta_{n_1+n_2} & \delta_{n_1+n_2} \\
(0,0) & \ldots & (0,0) & (1,1) & (0,0) & \ldots & (0,0) & (0,1)
\end{array}
\end{equation}
Again, we see that the root space decomposition of $\osp(1,0|2n_1,2n_2)$ is very similar to that of $\osp(1|2n_1+2n_2)$, but the 
degree of the roots is different, and of course the bracket between root vectors is also different.

\subsection{$\mathbf{\Z_2\times\Z_2}$-graded parabosons}

Ordinary parabosons are realized as the root vectors of the Lie superalgebra $\osp(1|2n)$, corresponding to the short roots.
Here, for $\g=\osp(1,0|2n_1,2n_2)$, we can realize $\Z_2\times\Z_2$-graded parabosons by means of the following root vectors  
(the factor $\sqrt{2}$ is again chosen for convenience):
\begin{equation}
b_{k}^-= \sqrt{2}(e_{1,k+1}-e_{n_1+n_2+k+1,1}),\quad b_{k}^+=\sqrt{2}(e_{1,n_1+n_2+k+1}+e_{k+1,1}),
\end{equation}
where $k=1,\ldots,n_1+n_2$.
The graded subspaces of $\g$ are:
\begin{align*}
& \g_{(1,0)}=\spn \{ b_{k}^{\pm},\; k=1,\ldots,n_1 \} \\
& \g_{(0,1)}=\spn \{b_{k}^{\pm},\; k=n_1+1,\ldots,n_1+n_2  \} \\
& \g_{(0,0)}=\spn \{ \{ b_{k}^\xi,  b_{l}^\eta \},\; \xi, \eta =\pm, \; k,l=1,\ldots,n_1 \; \hbox{and} \; k,l=n_1+1,\ldots,n_1+n_2   \} \\
& \g_{(1,1)}=\spn \{ [b_{k}^{\xi}, b_{l}^{\eta}], \; \xi, \eta =\pm, \;k= 1,\ldots ,n_1,\; l=n_1+1,\ldots n_1+n_2; \\
& \qquad\qquad\qquad \hbox{or } k=n_1+1,\ldots,n_1+n_2, \; l=1,\ldots,n_1 \}.
\end{align*}
In other words, we are dealing with two ensembles of parabosons: those of degree $(1,0)$ and those of degree $(0,1)$.
Within one of these sets, the common paraboson relations~\eqref{pb} are valid:
\begin{align}
& [\{ b_{ j}^{\xi}, b_{ k}^{\eta}\} , b_{l}^{\epsilon}]= 
(\epsilon -\xi) \delta_{jl} b_{k}^{\eta}  + (\epsilon -\eta) \delta_{kl}b_{j}^{\xi}, \nonumber\\
& \hbox{either }j,k,l\in \{1,2,\ldots,n_1\} \hbox{ or else } j,k,l\in \{n_1+1,\ldots,n_1+n_2\}.
\end{align}
However, between the two sets of parabosons, the ``relative commutation relations'' are of a different type:
\begin{align}
& \{ [b_{ j}^{\xi}, b_{ k}^{\eta}] , b_{l}^{\epsilon}\}= 
-(\epsilon -\xi) \delta_{jl} b_{k}^{\eta}  + (\epsilon -\eta) \delta_{kl}b_{j}^{\xi}, \nonumber \\
& \hbox{either } j=1,\ldots,n_1,\  k=n_1+1,\ldots,n_1+n_2, \nonumber\\
& \hbox{or else } j=n_1+1,\ldots,n_1+n_2,\ k=1,\ldots,n_1; \nonumber\\
& l=1,\ldots,n_1+n_2.
\label{relpf-2}
\end{align}
This describes a simultaneous system of two sets of parabosons with ``relative parafermion relations''~\eqref{relpf-2},
for which the underlying algebraic structure is a $\Z_2\times\Z_2$-graded Lie superalgebra.

\section{Conclusions}

Since their introduction by Green~\cite{Green}, parabosons and parafermions have continued to inspire physicists and mathematicians.
The relation of these parastatistics operators with Lie algebras and Lie superalgebras dates from decades ago.
But only recently it was realized that a system of parafermions and parabosons satisfying so-called relative paraboson relations is no longer governed by a Lie algebra or a Lie superalgebra, but by a $\Z_2\times\Z_2$-graded Lie superalgebra~\cite{Tolstoy2014}.
This raised the interest in $\Z_2\times\Z_2$-graded Lie algebras and  $\Z_2\times\Z_2$-graded Lie superalgebras.

In the current paper, we describe some classical $\Z_2\times\Z_2$-graded Lie algebras and superalgebras.
For the orthogonal $\Z_2\times\Z_2$-graded Lie algebra $\so_q(2n+1)$ we give a description of its roots and root vectors. 
These are now graded by an element of $\Z_2\times\Z_2$, in other words, there are four types of roots and root vectors.
As far as we know, this is the first time that some structure theory is developed for a $\Z_2\times\Z_2$-graded Lie algebra.
The algebra $\so_q(2n+1)$ is also relevant for parastatistics: if the creation and annihilation operators of parafermions are
identified with the short root vectors of $\so_q(2n+1)$, one obtains a system of two ensembles of parafermions.
Each ensemble itself satisfies the standard parafermion commutation relations.
But the relations mixing these two ensembles are in terms of nested anti-commutation relations.

In a similar way, we describe the roots and root vectors of the $\Z_2\times\Z_2$-graded Lie superalgebra $\osp(1,0|2n_1,2n_2)$.
Also this algebra is relevant for parastatistics: identifying the creation and annihilation operators of parabosons with
the short root vectors of $\osp(1,0|2n_1,2n_2)$ leads to a system of two ensembles of parabosons.
Each ensemble itself satisfies the standard triple relations of parabosons, but the relations mixing these two ensembles are 
triple relations with commutators replaced by anti-commutators and vice versa.

It should be clear that one can continue along these lines.
In particular, it should be possible to relate the general orthosymplectic $\Z_2\times\Z_2$-graded Lie superalgebra $\osp(2m_1+1,2m_2|2n_1,2n_2)$ described in~\cite{SV6} to a system consisting of two ensembles of parafermions and two ensembles of parabosons with proper mixed relations.

These examples show that the study of $\Z_2\times\Z_2$-graded Lie algebras and superalgebras lead to new and interesting parastatistics systems which are at first sight similar to the classical ones~\cite{Green,GM}, but which are fundamentally different.
The study of further properties of these new systems would be interesting, but requires representation theory (i.e.\ the construction of Fock spaces) of the algebras involved.
In general, the explicit construction of parastatistics Fock spaces is a very difficult and involved problem~\cite{SV7,SV8,SV1}.

\section*{Acknowledgments}
Both authors were supported by the Bulgarian National Science Fund, grant KP-06-N88/3.






\end{document}